\documentclass[]{emulateapj}
\begin{document}

\title{The Sizes of Early-type Galaxies}			
\author{Joachim Janz \& Thorsten Lisker}
\affil{Zentrum f\"ur Astronomie der Universit\"at Heidelberg (ZAH), M\"onchhofstra{\ss}e 12-14, D-69120 Heidelberg, Germany}
\email{jjanz@ari.uni-heidelberg.de}

\submitted{}

\received{\textit{1st September 2008}}
\revised{\textit{14th October 2008}}
\accepted{\textit{15th October 2008}}

\begin{abstract}
In this letter we present a study of the size luminosity relation of 475 early-type galaxies in the Virgo Cluster with Sloan Digital Sky Survey imaging data. The analysis of our homogeneous, model-independent data set reveals that giant and dwarf early-type galaxies do not form one common sequence in this relation. The dwarfs seem to show weak or no dependence on luminosity, and do not fall on the extension of the  rather steep relation of the giants.
Under the assumption that the light profile shape varies continuously with magnitude, a curved relation of
size and magnitude would be expected. While the galaxies do roughly follow this trend overall, we find that the dwarf galaxies are significantly larger 
and the low-luminosity giants are significantly smaller than what is predicted. 
We come to the conclusion that in this scaling relation there is not one common sequence from dwarfs to giants, but a dichotomy which can not be resolved by varying profile shapes. The comparison of our data to a semi-analytic model supports the idea of a physical origin of this dichotomy.
\end{abstract}


\section{Introduction}
Scaling relations have ever been an important tool not only to study galaxy properties but also to link those properties to their formation and evolution. Since early-type galaxies are the most numerous galaxy type in cluster environments, they play an outstanding  role in understanding galaxy clusters and therefore structure formation in general. But still today it remains an open question whether the giant early-type galaxies and their lighter counterparts share the same origin and formation mechanisms. Of the morphological scaling relations, so far mostly the relation between surface brightness and size  \citep[``Kormendy relation", ][]{1985ApJ...295...73K}, and between surface brightness and luminosity (e.g. \citealt{binggeli_cameron})  were studied to tackle this question. Combined with velocity dispersion, the Faber-Jackson relation \citep{1976ApJ...204..668F} and the extension to the Fundamental Plane \citep{1987ApJ...313...42D,1987ApJ...313...59D} were studied, mostly for giant early types. But even with the now available facilities velocity dispersions are still rare for dwarf galaxies.

Until the early 1990's sizes of nearby early-type galaxies were studied for example by \citet{1977ApJ...218..333K}, \citet{1993MNRAS.265..731G}, and for the Virgo Cluster in particular by \citet{binggeli_cameron} for dwarfs and by \citet{1993MNRAS.265.1013C}  for giants. An often cited source of a homogenous data set of sizes for dynamically hot systems over the whole luminosity range is \citet{1992ApJ...399..462B}. The conclusion of that time was that giant and dwarf early-type galaxies show a distinct size distribution, and that the dwarfs show less change of size with luminosity than the giants. This together with other scaling relations was interpreted as evidence for a different origin of dwarf and giant early-type galaxies.

Towards the turn of the millenium, however, it became more widely realized that the light profile shapes of early types vary continuously with luminosity. Neither do dwarf galaxies simply follow exponential profiles, nor do all giants exhibit de Vaucouleurs profiles. Instead, all early types are well described by the generalized S\'ersic profile \citep{1963BAAA....6...41S} with different S\'ersic indices $n$ \citep{1994MNRAS.268L..11Y,2006ApJS..164..334F}. Several authors reasoned that the scaling relations naturally follow what is predicted by $n$ changing linearly with magnitude, and that all these galaxies can indeed be of the same kind \citep{jerjen_binggeli,1998A&A...333...17B,graham_guzman,2005A&A...430..411G}.
While this latter interpretation is, for example, not shared by
\citet{2008arXiv0807.3282B}, who argue that dwarf and giant early-type
galaxies have different origins, these authors do agree that they can
be seen as one structural family with a gradual variation of $n$ with
luminosity. 
Recently, \citet{kormendy08} conclude that dwarfs and giants are \textit{structurally}
distinct, notwithstanding the S\'ersic continuum, which they consider insensitive to the physics dividing the two.
In our analysis below we quantitatively pin down this structural distinction by taking the variation of profile shapes explicitly into account.

\section{Sample Selection and Imaging Data}

Our sample is selected based on the Virgo Cluster Catalog (VCC; \citealt{b_s_t}).
Only
certain cluster members with $m_B<18.0$ mag are taken into account,
which is the same magnitude limit up to which the VCC was found to be complete.  
This translates into $M_B<-13.0$ mag with our adopted distance modulus of m-M=31.0 mag (d=15.85 Mpc, \citealt{1999ApJ...516..626G}).

Galaxies listed as 
``S0:", ``E/S0'', ``S0/Sa'', and ``SB0/SBa'' are taken as S0, and
one S0 (VCC1902) is excluded, since it shows clear spiral arm
structure. For the dwarfs, we selected galaxies classified as
dE, dS0, and ``dE:", whereas ``dE/Im'' as well as possible irregulars
based on visual inspection are excluded \citep{lisker_etal}. 
We excluded 13 dwarfs where the Petrosian aperture (see below) could not be obtained, as well as 
13 dwarfs and four giants that are too strongly contaminated by the
light of close neighbour objects. This leads to a working sample of 475 galaxies: 397 early-type
dwarfs (``dEs"), 9 M32-type candidates \citep[as listed in Table XIII
of][]{b_s_t}, and 67 E and S0 galaxies.

\label{sec:imagingdata}

The Sloan Digital Sky Survey (SDSS)
Data Release Five \citep{2007ApJS..172..634A} covers all but six
early-type dwarf galaxies of the VCC. The pixel scale of  $0''\!\!\!.396$ corresponds to a physical size of 30 pc.
For the analysis below, we use the
\textit{r}-band, which has the highest S/N.
Since the quality of sky level subtraction of the SDSS pipeline 
is insufficient, we use sky-subtracted images as provided by
\citet{lisker_etal}, based on a careful subtraction method.
The images were flux-calibrated and corrected for Galactic extinction
\citep{1998ApJ...500..525S}.  

The image analysis is done largely in the same way as in
\citet{2008AJ....135..380L}: 
For each galaxy, we determined a ``Petrosian semimajor axis"  (\citealt{1976ApJ...209L...1P}, $a_p$ ), i.e., we
use ellipses instead of circles in the calculation of the Petrosian  
radius (see, e.g., \citealt{2004AJ....128..163L}). The total flux in the
$r$ band was measured within $a = 2 a_p$ , yielding a value for 
the half-light semimajor axis, $a_{hl,r,uncorr}$. 
This Petrosian aperture still misses some flux, which is of particular
relevance for the giant galaxies \citep{2001MNRAS.326..869T}. As improvement of
\citet{2008AJ....135..380L}, luminosities and half light radii are
corrected for this missing flux according to
\citet{2005AJ....130.1535G}. Axial ratio and position angle were then
determined through an isophotal fit at  $a = 2 a_{hl,r}$. The
effective radius is then given by
$r_{\textit{eff}}=a_{\textit{hl,r}}\sqrt{b/a}$ with the axis ratio
$b/a$.  
Additionally we fitted S\'ersic profiles. 
We omitted intensities at radii $r <
2^{\prime\prime}$ in order to avoid seeing effects.

Our data set is a very homogeneous set of parameters for
galaxies in one cluster, based on 
data taken with the same instrument and reduced and analyzed with the
same procedure. We point out that our derived radii are model
independent.\footnote{The corrections for the missing flux are based on the
concentration parameter of the galaxies; therefore they implicitely
depend on the assumption of a (S\'ersic) light profile model
\citep[cf.][]{2005AJ....130.1535G}.}

\section{Sizes of early-type galaxies}
 In Fig.~1 we present the size luminosity diagram for our sample. At
first glance the sequence from dwarf to giant early-type galaxies does
not look very continuous: the giants follow a steep relation with a
well-defined edge on the bright end of their distribution. The bunch of
dwarfs apparently lie with a larger scatter around an effective
radius of $r_{\textit{eff}}=1$ kpc, their sizes showing weak to no
dependence on luminosity.\footnote{The almost constancy of the dwarfs
  is just a coincidence when taking the half light radius containing
  50~\% of the light. For example, for the
  radius containing 90~\% of the light, the relation for
  dwarfs steepens. This can be understood when taking the gradual
  variation of profile shapes with luminosity (see below) into
  account.}{} 

Without applying the correction for the missing flux within the Petrosian
aperture, the separation between the two sequences even widens (not shown) since the
less compact objects of the same luminosity show less concentration and
therefore a smaller correction. 

It is important to notice that the different behaviour of dwarfs and
giants does not depend on whether objects with
 disk components are omitted. This can be seen from Fig.~2, where
gray symbols indicate objects with disk components or disk-like
structure.

Instead of adopting one distance for all galaxies, the Virgo cluster
can alternatively be described as a substructured system, with the
different components partly having different distances
\citep{1999MNRAS.304..595G}. Since this can affect the apparent size of
a galaxy, we checked whether the distribution of galaxies within the
size luminosity diagram correlates with projected position in the cluster.
We do not find any such correlation. Therefore, while distance
variations could explain the larger
scatter of the sizes of early types in the Virgo Cluster as compared
to other clusters \citep{2008arXiv0807.3282B}, it cannot explain the observed dichotomy: \citeauthor{2008arXiv0807.3282B}'s Fig.~6 indicates the apparent separation of
dwarfs and giants and they use different distance moduli for different
subparts of the Virgo cluster.

The same impression can also be obtained from other previous studies,
e.g.\ \citet[]{binggeli_cameron}, Fig.~1b;
\citet[]{1992ApJ...399..462B}, Table~1; \citet[]{kormendy08}, Fig.~37. It is, however, not as clearly
seen in the compilation of sizes of elliptical galaxies from several
different studies presented by \citet{2008MNRAS.tmp..752G} (Fig.\ 10). In this
more heterogeneous dataset, the relative number of small
low-luminosity giants as well as that of large bright dwarfs appears to be
somewhat smaller.

\begin{figure}
\includegraphics[scale=0.85,angle=0]{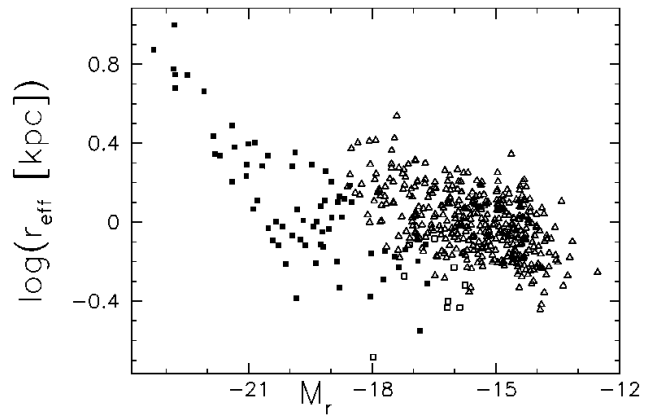}
\caption{Absolute magnitude in $r$ versus logarithm of half light radius. dEs
  are shown with open triangles  and E and S0 with filled squares. M32-type
  candidates are drawn with open squares.} 
\end{figure}

\begin{figure*}
\includegraphics[scale=.85,angle=0]{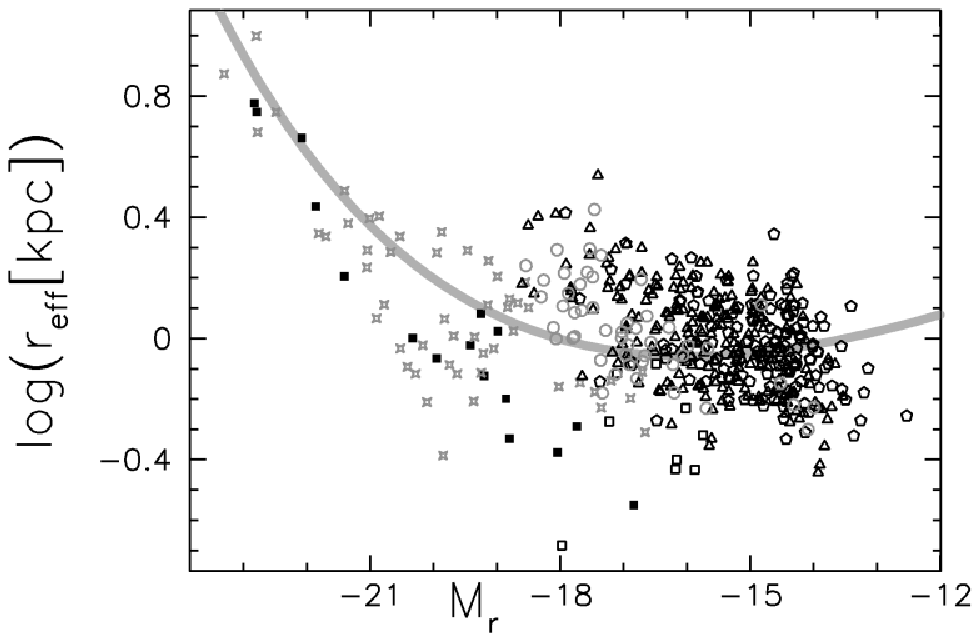}\hfill
\includegraphics[scale=.85,angle=0]{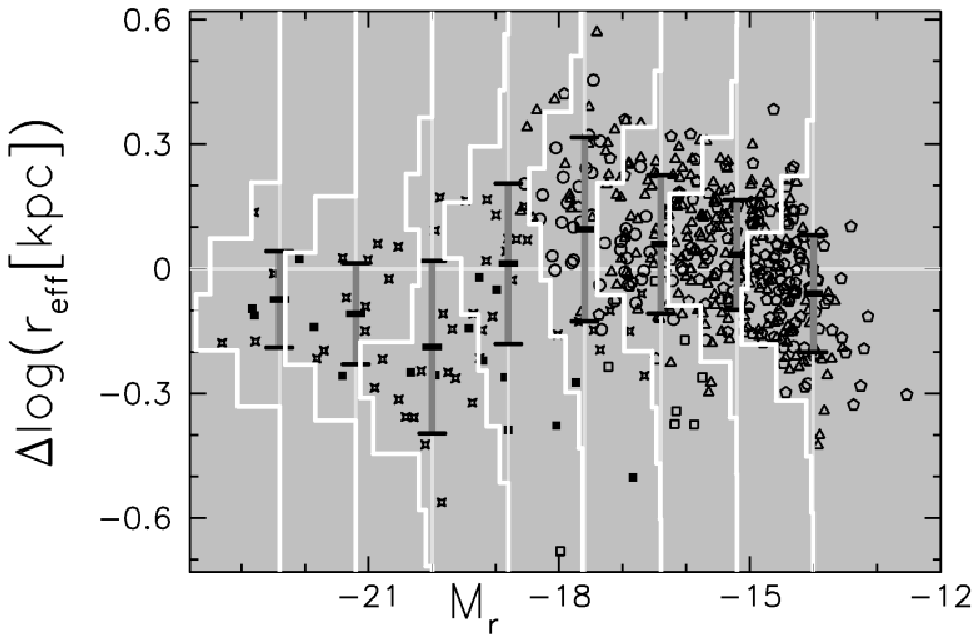}\hfill
\caption{\emph{Left panel:} Same as Fig.~1, with prediction for
  S\'ersic profiles with the S\'ersic index $n$ given as a function of
  luminosity. Now the sub-classifications according to
  \citet{lisker_etal} are taken into account. The symbols mean: filled
  squares - E, gray stars - S0, open squares - M32 candidates, open
  triangles - dE,N, open pentagons - dE,nN and gray open circles for
  dwarf galaxies with probable disk-like structure (dE(di) or dE(bc)). 
\emph{Right panel:} The residuals in $\log(r)$ about the predicted
relation.  The vertical
histograms show the galaxy distributions in magnitude bins of 1.2
mag. Their zero levels are indicated by the thin gray lines.
Thick gray lines measure the scatter around the mean (black marks),
with the standard deviation indicated in black at the ends of the
lines.} 
\end{figure*}

\subsection{Varying profile shapes ?}
\citet{graham_guzman} suggested that the apparent dichotomy between
dwarfs and giants in scaling relations can be explained just by the
fact that the profile shape of a galaxy scales with magnitude. They
describe the light profiles with S\'ersic profiles and show the effect
of a linear relation between magnitude and logarithm of the S\'ersic
index $n$ on the other scaling relations. As a result, the dependence of effective
radius on magnitude becomes stronger at higher luminosities and the brightest
galaxies are naturally larger (Fig.~11 in
\citealt{2008MNRAS.tmp..752G}).

For investigating whether our galaxies 
display the predicted behavior, we fitted their azimuthally averaged light
profiles with S\'ersic models (see Sect.~\ref{sec:imagingdata}). The derived
S\'ersic indices $n$ and central surface brightnesses $\mu_0$ were
then used to obtain linear fits to the $\mu_0/M_r$ and $n/M_r$
relations, using a least squares fitting algorithm.  For those fits we
exclude systems with a (probable) disk component, namely galaxies
classified as S0, dEs with disk features, 
 and dEs
with blue centers \citep{2006AJ....132..497L,2006AJ....132.2432L}. This is to ensure that the
light profiles can be well parametrized by S\'ersic profiles.
 Our fits together with equation (16) of
\citet{2008MNRAS.tmp..752G} predict a non-linear sequence in
the $r_{\textit{eff}}/M_r$ diagram.

The predicted relation is shown together with the observed galaxies
in the left panel of Fig.~2. With the visual guidance of the line,
it appears more likely that the data points follow one common continuous
relation. And the gross trend in the diagram can indeed be explained
by varying profile shapes. However, at
luminosities brightwards of the transition between dwarfs and giants, a
substantial amount of galaxies fall below the relation, while
faintwards most of the dwarfs lie above it.

We note that M32-type candidates play a minor role just by their small
number, and furthermore, they are fainter than the ``compact giants" in
question. Whether both of these 
are special enough to justify their own classification is out of
the scope of this letter, and will be investigated in a future study.

To quantify the
departure of the observed galaxies from the predicted relation we show
in the right panel of Fig.~2 the size residuals about the curve
shown in the left panel. While at both ends of the luminosity range
the observed galaxies  fall onto the relation,
 it can now be seen even more clearly that towards intermediate
 luminosities, they depart more and more from the
relation in opposite directions. Brightwards of the transition region,
most galaxies are only half the predicted size, while faintwards, many
galaxies are larger than predicted by $\ge$50\%.

Furthermore, we divide the data into magnitude bins and
investigate the distribution of galaxies in those bins with histograms
(Fig.~2, right panel). The scatter and the shapes of
the histograms confirm the coexistence of two separate relations. In the
two bins in which the transition occurs, the scatter increases, and is
even larger than for the faintest galaxies, for which one would
naively expect the largest scatter. Moreover,
the distribution around the mean changes from the Gaussian-like shape
seen at the bright and faint end to a much broader, even double-peaked
shape. For those two bins, a K-S test yields a probability for the
null hypothesis that the two samples follow the same distribution of
$2.11\%$, which means that the break is statistically significant. 

Only the bright end of the predicted curve shows strong sensitivity to
the fitted relations. At intermediate luminosities, it is
stable against small changes of the
fits. Moreover, a change of the curve would not reduce the significant
difference between the bright dwarfs and
low-luminosity giants.

Thus our analysis shows two things. First, the residuals do not
resemble a quite large random scatter around the relation. And
therefore the size luminosity relation can not be fully explained by
varying profile shapes. Second, the abrupt change in the behavior of
faint and bright galaxies is even emphasized through the above
examination,  and this break is a real
discontinuity of the sequence from lowest to highest luminosities.

\section{Comparison to Semi-Analytic Model}
\label{sec:sam}

The Numerical Galaxy Catalog of \cite{2005ApJ...634...26N} is
based on a high resolution $N$-body simulation in a 
$\Lambda$CDM
 universe
\citep{2005PASJ...57..779Y}.  The dark-halo merger trees of the
$N$-body simulation are taken as input for a semi-analytic model
of the physical processes governing galaxy formation and evolution (here a modified 
version of the Mitaka Model, 
\citealt{2004ApJ...610...23N}).

In particular, this model takes into account the dynamical response
to starburst-induced gas removal after gas-rich mergers (also for cases intermediate
between a purely baryonic cloud and a baryonic cloud fully supported by
surrounding dark matter as in \citealt{1987A&A...188...13Y}). 
This process plays a crucial role for the sizes of early-type dwarf
galaxies. 
Their gravitational well is shallower and
thus they suffer a more substantial gas loss than giants. The subsequent variation
of the potential results in an increase in size. If it is not taken
into account the dwarf galaxies are modelled to be systematically
smaller.

\begin{figure}
\includegraphics[scale=.85,angle=0]{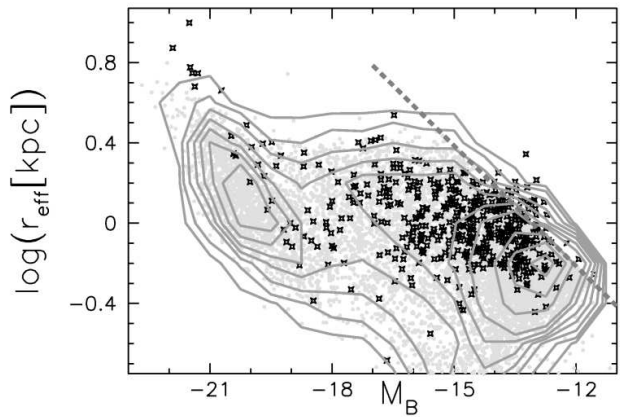}
\caption{Absolute $B$ magnitude versus logarithm of half light
  radius for observed (black) and model
  galaxies (gray) from \cite{2005ApJ...634...26N}. Model galaxies with fainter surface brightnesses than a limit of
    $\left<\mu_{B}\right><25.5\ \textrm{mag/arcsec}^2$ (dashed line)
 are excluded. Contours were calculated using model galaxy abundances in bins of 0.75 mag and 0.2 dex. The contour levels, relative to the lowest, are 2.5, 5, 8, 11, 15, 20, 35, 50} 
\end{figure}

We identify model galaxies as
early-type galaxies if they are bulge dominated (bulge to total ratio $>0.6)$.
To compare our data with the model we transformed SDSS $g$ magnitudes into $B$ according to \citet{2002AJ....123.2121S},
 using the galaxies' $g-r$ color measured
within $a_{hl,r}$. In Fig.~3 one can see that the model
galaxies show a bimodality similar to what we observe, with low galaxy
density between the two regions. 
In those dwarfs which form by gas-rich major mergers a starburst follows and the dwarfs are enlarged by the dynamic response to the subsequent gas loss. This mechanism is not at work in gas-deficient mergers, and the resulting galaxies are smaller.

Note that \citeauthor{2005ApJ...634...26N} assume de Vaucouleurs
profiles to calculate projected half-light radii from half-mass radii.
For exponential profiles, which would be more appropriate for dwarfs, 
the model galaxies would  shift upwards in the diagram by 0.11 dex
\citep{2003MNRAS.340..509N}.

It is interesting, that they compared their model to the data of \citet{1992ApJ...399..462B} and
concluded
that a substantial fraction of model dwarfs is too large to match the observed galaxies.
With our dataset, a much better agreement is found (also see \citealt{2005A&A...438..491D}).

\section{Summary and Discussion}

We studied the size luminosity relation of Virgo cluster early-type
galaxies, based on model-independent size measurements from SDSS
imaging data. We find that the dwarfs 
do not fall on the extension of the  rather steep sequence of the giants.
While the gross trend in the size luminosity relation can be
explained by light profile shapes becoming steeper for more luminous
galaxies, a closer look reveals that there is a clear discontinuity in
the behaviour of faint and bright galaxies.

\citet{2008MNRAS.386..864D} compare dynamical masses and projected
half-light radii of elliptical galaxies, bulges of spiral galaxies,
dwarf spheroidals, massive compact objects (MCOs), and globular clusters
(GCs). These authors also find a bimodality in the size distribution of
elliptical galaxies. The ``compact low-mass ellipticals" and the 
giant ellipticals lie on one sequence down to the MCOs and GCs, while
the ellipticals of intermediate brightness in the larger branch lie on one sequence with the dwarf spheroidals. 
 The interpretation of \citet{2000ApJ...543..149O}, based on
 calculations of galaxy interactions,  is that those latter galaxies
are mostly of tidal
origin and that  the ``compact low-mass ellipticals" should be
interpreted as the real counterparts of the more massive elliptical
galaxies. 
This is an alternative interpretation to the one based on semi-analytic models (Sect.~\ref{sec:sam}), and represents further
evidence for the physical relevance of the
observed dichotomy. 

Our findings are a new piece to the
puzzle of the connection between dwarf and giant early-type galaxies
and may hint at different origins. 
A further analysis of the scaling relations and comparison with models
seems a promising approach to bring more light to this
question.
In particular it will be interesting to investigate correlations with
velocity dispersion, since it should be closely related to size evolution
in the framework of dynamical response.
We will pursue this idea in a forthcoming paper.

\acknowledgements
We thank Masahiro Nagashima for helpful advice.
   T.L.\ is supported by the Excellence Initiative
    whithin the German Research Foundation (DFG).
The study is based on SDSS (http://www.sdss.org/). 


\clearpage


\begin{thebibliography}{42}
\expandafter\ifx\csname natexlab\endcsname\relax\def\natexlab#1{#1}\fi

\bibitem[{{Adelman-McCarthy} {et~al.}(2007){Adelman-McCarthy}, {Ag{\"u}eros},
  {Allam}, {Anderson}, {Anderson}, {Annis}, {Bahcall}, {Bailer-Jones},
  {Baldry}, {Barentine}, {Beers}, {Belokurov}, {Berlind}, {Bernardi},
  {Blanton}, {Bochanski}, {Boroski}, {Bramich}, {Brewington}, {Brinchmann},
  {Brinkmann}, {Brunner}, {Budav{\'a}ri}, {Carey}, {Carliles}, {Carr},
  {Castander}, {Connolly}, {Cool}, {Cunha}, {Csabai}, {Dalcanton}, {Doi},
  {Eisenstein}, {Evans}, {Evans}, {Fan}, {Finkbeiner}, {Friedman}, {Frieman},
  {Fukugita}, {Gillespie}, {Gilmore}, {Glazebrook}, {Gray}, {Grebel}, {Gunn},
  {de Haas}, {Hall}, {Harvanek}, {Hawley}, {Hayes}, {Heckman}, {Hendry},
  {Hennessy}, {Hindsley}, {Hirata}, {Hogan}, {Hogg}, {Holtzman}, {Ichikawa},
  {Ichikawa}, {Ivezi{\'c}}, {Jester}, {Johnston}, {Jorgensen}, {Juri{\'c}},
  {Kauffmann}, {Kent}, {Kleinman}, {Knapp}, {Kniazev}, {Kron}, {Krzesinski},
  {Kuropatkin}, {Lamb}, {Lampeitl}, {Lee}, {Leger}, {Lima}, {Lin}, {Long},
  {Loveday}, {Lupton}, {Mandelbaum}, {Margon}, {Mart{\'{\i}}nez-Delgado},
  {Matsubara}, {McGehee}, {McKay}, {Meiksin}, {Munn}, {Nakajima}, {Nash},
  {Neilsen}, {Newberg}, {Nichol}, {Nieto-Santisteban}, {Nitta}, {Oyaizu},
  {Okamura}, {Ostriker}, {Padmanabhan}, {Park}, {Peoples}, {Pier}, {Pope},
  {Pourbaix}, {Quinn}, {Raddick}, {Re Fiorentin}, {Richards}, {Richmond},
  {Rix}, {Rockosi}, {Schlegel}, {Schneider}, {Scranton}, {Seljak}, {Sheldon},
  {Shimasaku}, {Silvestri}, {Smith}, {Smol{\v c}i{\'c}}, {Snedden}, {Stebbins},
  {Stoughton}, {Strauss}, {SubbaRao}, {Suto}, {Szalay}, {Szapudi}, {Szkody},
  {Tegmark}, {Thakar}, {Tremonti}, {Tucker}, {Uomoto}, {Vanden Berk},
  {Vandenberg}, {Vidrih}, {Vogeley}, {Voges}, {Vogt}, {Weinberg}, {West},
  {White}, {Wilhite}, {Yanny}, {Yocum}, {York}, {Zehavi}, {Zibetti}, \&
  {Zucker}}]{2007ApJS..172..634A}
{Adelman-McCarthy}, J.~K., et.~al. 2007, \apjs, 172, 634

\bibitem[{{Bender} {et~al.}(1992){Bender}, {Burstein}, \&
  {Faber}}]{1992ApJ...399..462B}
{Bender}, R., {Burstein}, D., \& {Faber}, S.~M. 1992, \apj, 399, 462

\bibitem[{{Binggeli} \& {Cameron}(1991)}]{binggeli_cameron}
{Binggeli}, B. \& {Cameron}, L.~M. 1991, \aap, 252, 27

\bibitem[{{Binggeli} \& {Jerjen}(1998)}]{1998A&A...333...17B}
{Binggeli}, B. \& {Jerjen}, H. 1998, \aap, 333, 17

\bibitem[{{Binggeli} {et~al.}(1985){Binggeli}, {Sandage}, \& {Tammann}}]{b_s_t}
{Binggeli}, B., {Sandage}, A., \& {Tammann}, G.~A. 1985, \aj, 90, 1681

\bibitem[{{Boselli} {et~al.}(2008){Boselli}, {Boissier}, {Cortese}, \&
  {Gavazzi}}]{2008arXiv0807.3282B}
{Boselli}, A., {Boissier}, S., {Cortese}, L., \& {Gavazzi}, G. 2008, \aap,  in press

\bibitem[{{Caon} {et~al.}(1993){Caon}, {Capaccioli}, \&
  {D'Onofrio}}]{1993MNRAS.265.1013C}
{Caon}, N., {Capaccioli}, M., \& {D'Onofrio}, M. 1993, \mnras, 265, 1013


\bibitem[{{Dabringhausen} {et~al.}(2008){Dabringhausen}, {Hilker}, \&
  {Kroupa}}]{2008MNRAS.386..864D}
{Dabringhausen}, J., {Hilker}, M., \& {Kroupa}, P. 2008, \mnras, 386, 864

\bibitem[{{de Rijcke} {et~al.}(2005){de Rijcke}, {Michielsen}, {Dejonghe},
  {Zeilinger}, \& {Hau}}]{2005A&A...438..491D}
{de Rijcke}, S., {Michielsen}, D., {Dejonghe}, H., {Zeilinger}, W.~W., \&
  {Hau}, G.~K.~T. 2005, \aap, 438, 491

\bibitem[{{Djorgovski} \& {Davis}(1987)}]{1987ApJ...313...59D}
{Djorgovski}, S. \& {Davis}, M. 1987, \apj, 313, 59

\bibitem[{{Dressler} {et~al.}(1987){Dressler}, {Lynden-Bell}, {Burstein},
  {Davies}, {Faber}, {Terlevich}, \& {Wegner}}]{1987ApJ...313...42D}
{Dressler}, A., et.al. 1987, \apj, 313, 42

\bibitem[{{Faber} \& {Jackson}(1976)}]{1976ApJ...204..668F}
{Faber}, S.~M. \& {Jackson}, R.~E. 1976, \apj, 204, 668

\bibitem[{{Ferrarese} {et~al.}(2006){Ferrarese}, {C{\^o}t{\'e}}, {Jord{\'a}n},
  {Peng}, {Blakeslee}, {Piatek}, {Mei}, {Merritt}, {Milosavljevi{\'c}},
  {Tonry}, \& {West}}]{2006ApJS..164..334F}
{Ferrarese}, L., et.~al. 2006, \apjs, 164,
  334

\bibitem[{{Gavazzi} {et~al.}(1999){Gavazzi}, {Boselli}, {Scodeggio}, {Pierini},
  \& {Belsole}}]{1999MNRAS.304..595G}
{Gavazzi}, G., {Boselli}, A., {Scodeggio}, M., {Pierini}, D., \& {Belsole}, E.
  1999, \mnras, 304, 595

\bibitem[{{Gavazzi} {et~al.}(2005){Gavazzi}, {Donati}, {Cucciati}, {Sabatini},
  {Boselli}, {Davies}, \& {Zibetti}}]{2005A&A...430..411G}
{Gavazzi}, G., {Donati}, A., {Cucciati}, O., {Sabatini}, S., {Boselli}, A.,
  {Davies}, J., \& {Zibetti}, S. 2005, \aap, 430, 411

\bibitem[{{Graham} {et~al.}(2005){Graham}, {Driver}, {Petrosian}, {Conselice},
  {Bershady}, {Crawford}, \& {Goto}}]{2005AJ....130.1535G}
{Graham}, A.~W., et.~al. 2005, \aj, 130, 1535

\bibitem[{{Graham} \& {Guzm{\'a}n}(2003)}]{graham_guzman}
{Graham}, A.~W. \& {Guzm{\'a}n}, R. 2003, \aj, 125, 2936

\bibitem[{{Graham} \& {Worley}(2008)}]{2008MNRAS.tmp..752G}
{Graham}, A.~W. \& {Worley}, C.~C. 2008, \mnras, 752

\bibitem[{{Graham} {et~al.}(1999){Graham}, {Ferrarese}, {Freedman},
  {Kennicutt}, {Mould}, {Saha}, {Stetson}, {Madore}, {Bresolin}, {Ford},
  {Gibson}, {Han}, {Hoessel}, {Huchra}, {Hughes}, {Illingworth}, {Kelson},
  {Macri}, {Phelps}, {Sakai}, {Silbermann}, \& {Turner}}]{1999ApJ...516..626G}
{Graham}, et.~al.1999, \apj,
  516, 626

\bibitem[{{Guzman} {et~al.}(1993){Guzman}, {Lucey}, \&
  {Bower}}]{1993MNRAS.265..731G}
{Guzman}, R., {Lucey}, J.~R., \& {Bower}, R.~G. 1993, \mnras, 265, 731

\bibitem[{{Jerjen} \& {Binggeli}(1997)}]{jerjen_binggeli}
{Jerjen}, H. \& {Binggeli}, B. 1997, in ASP Conf. Ser. 116, The Nature of Elliptical Galaxies; ed. M.~{Arnaboldi}, G.~S. {Da Costa}, \& P.~{Saha}, 239

\bibitem[{{Kormendy}(1977)}]{1977ApJ...218..333K}
{Kormendy}, J. 1977, \apj, 218, 333

\bibitem[{{Kormendy}(1985)}]{1985ApJ...295...73K}
---. 1985, \apj, 295, 73

\bibitem[{{Kormendy} {et~al.}(2008){Kormendy}, {Fisher}, {Cornell}, \&
  {Bender}}]{kormendy08}
  {Kormendy}, J., {Fisher}, D.~B., {Cornell}, M.~E., \& {Bender}, R. 2008, \apjs, in press

\bibitem[{{Lisker} {et~al.}(2006{\natexlab{a}}){Lisker}, {Glatt}, {Westera}, \&
  {Grebel}}]{2006AJ....132.2432L}
{Lisker}, T., {Glatt}, K., {Westera}, P., \& {Grebel}, E.~K.
  2006{\natexlab{a}}, \aj, 132, 2432

\bibitem[{{Lisker} {et~al.}(2006{\natexlab{b}}){Lisker}, {Grebel}, \&
  {Binggeli}}]{2006AJ....132..497L}
{Lisker}, T., {Grebel}, E.~K., \& {Binggeli}, B. 2006{\natexlab{b}}, \aj, 132,
  497

\bibitem[{{Lisker} {et~al.}(2008){Lisker}, {Grebel}, \&
  {Binggeli}}]{2008AJ....135..380L}
---. 2008, \aj, 135, 380

\bibitem[{{Lisker} {et~al.}(2007){Lisker}, {Grebel}, {Binggeli}, \&
  {Glatt}}]{lisker_etal}
{Lisker}, T., {Grebel}, E.~K., {Binggeli}, B., \& {Glatt}, K. 2007, \apj, 660,
  1186

\bibitem[{{Lotz} {et~al.}(2004){Lotz}, {Primack}, \&
  {Madau}}]{2004AJ....128..163L}
{Lotz}, J.~M., {Primack}, J., \& {Madau}, P. 2004, \aj, 128, 163

\bibitem[{{Nagashima} {et~al.}(2005){Nagashima}, {Yahagi}, {Enoki}, {Yoshii},
  \& {Gouda}}]{2005ApJ...634...26N}
{Nagashima}, M., {Yahagi}, H., {Enoki}, M., {Yoshii}, Y., \& {Gouda}, N. 2005,
  \apj, 634, 26

\bibitem[{{Nagashima} \& {Yoshii}(2003)}]{2003MNRAS.340..509N}
{Nagashima}, M. \& {Yoshii}, Y. 2003, \mnras, 340, 509

\bibitem[{{Nagashima} \& {Yoshii}(2004)}]{2004ApJ...610...23N}
---. 2004, \apj, 610, 23

\bibitem[{{Okazaki} \& {Taniguchi}(2000)}]{2000ApJ...543..149O}
{Okazaki}, T. \& {Taniguchi}, Y. 2000, \apj, 543, 149

\bibitem[{{Petrosian}(1976)}]{1976ApJ...209L...1P}
{Petrosian}, V. 1976, \apjl, 209, L1

\bibitem[{{Schlegel} {et~al.}(1998){Schlegel}, {Finkbeiner}, \&
  {Davis}}]{1998ApJ...500..525S}
{Schlegel}, D.~J., {Finkbeiner}, D.~P., \& {Davis}, M. 1998, \apj, 500, 525

\bibitem[{{S{\'e}rsic}(1963)}]{1963BAAA....6...41S}
{S{\'e}rsic}, J.~L. 1963, Boletin de la Asociacion Argentina de Astronomia La
  Plata Argentina, 6, 41

\bibitem[{{Smith} {et~al.}(2002){Smith}, {Tucker}, {Kent}, {Richmond},
  {Fukugita}, {Ichikawa}, {Ichikawa}, {Jorgensen}, {Uomoto}, {Gunn}, {Hamabe},
  {Watanabe}, {Tolea}, {Henden}, {Annis}, {Pier}, {McKay}, {Brinkmann}, {Chen},
  {Holtzman}, {Shimasaku}, \& {York}}]{2002AJ....123.2121S}
{Smith}, et.~al. 2002, \aj, 123, 2121

\bibitem[{{Trujillo} {et~al.}(2001){Trujillo}, {Graham}, \&
  {Caon}}]{2001MNRAS.326..869T}
{Trujillo}, I., {Graham}, A.~W., \& {Caon}, N. 2001, \mnras, 326, 869


\bibitem[{{Yahagi}(2005)}]{2005PASJ...57..779Y}
{Yahagi}, H. 2005, \pasj, 57, 779

\bibitem[{{Yoshii} \& {Arimoto}(1987)}]{1987A&A...188...13Y}
{Yoshii}, Y. \& {Arimoto}, N. 1987, \aap, 188, 13

\bibitem[{{Young} \& {Currie}(1994)}]{1994MNRAS.268L..11Y}
{Young}, C.~K. \& {Currie}, M.~J. 1994, \mnras, 268, L11

\end{thebibliography}
\end{document}